\documentclass[12pt,prb]{revtex4}

\usepackage[english]{babel}
\usepackage{amsmath}
\usepackage[dvips]{graphicx}

\begin{document}

\begin{titlepage}

\title{ Lattice Models of Ionic Systems }

\author{Vladimir Kobelev, Anatoly B. Kolomeisky}
\affiliation{Department of Chemistry, Rice University, Houston, Texas 77005}

\author{\quad}
\author{Michael E. Fisher}
\affiliation{Institute for Physical Science and Technology, University of Maryland, College Park, Maryland 20742}

\begin{abstract}

A theoretical analysis of Coulomb systems on lattices in general dimensions is presented. The thermodynamics is developed  using Debye-H\"uckel theory with ion-pairing and dipole-ion solvation, specific calculations being performed for 3D lattices. As for continuum electrolytes, low-density results for  sc, bcc and fcc lattices indicate the existence of gas-liquid phase separation. The predicted critical densities have values comparable to those of continuum ionic systems, while the critical temperatures are 60-70\% higher. However, when the possibility of sublattice ordering as well as Debye screening is taken into account systematically, order-disorder transitions and a tricritical point are found on sc and bcc lattices, and gas-liquid coexistence is suppressed. Our results agree with recent Monte Carlo simulations of lattice electrolytes.
\end{abstract}

\maketitle

\end{titlepage}

\section{Introduction}

It is well  known  that criticality in simple fluids with short-range potentials  can be described by  the Ising universality class with critical exponents accessible via renormalization group calculations \cite{fisher74}. However, for Coulomb  systems, where particles interact through  long-range potentials,  the nature of criticality remains open to question \cite{fisher94}. Numerous theoretical, experimental and computational investigations of electrolyte systems have not yet produced a clear picture of the  thermodynamics in the critical region. Early experiments on criticality in electrolytes suggested a strong dichotomy: namely, some electrolytes \cite{japas90,narayanan94}, termed solvophobic and typically having large solvent dielectric constant, are satisfactorily characterized by Ising critical exponents. This suggests, that the principal interactions driving the phase separation in such systems are of short-range character. On the other hand, a number of organic salts in appropriate solvents, typically of low dielectric constant, were found to exhibit classical or close-to-classical behavior \cite{pitzer90}, and have been called Coulombic, stressing the importance of the dominant electrostatic interactions. Moreover, in sodium-ammonia solutions \cite{chieux70} (and some other systems: see \cite{bianchi01,weingartner01} and references therein), crossover from classical to Ising behavior had been observed, but at a reduced temperature $t\equiv(T-T_c)/T_c \simeq 0.6 \times 10^{-2}$, unusually close to the critical point. This led to the idea \cite{fisher94} that the true asymptotic critical behavior of ionic fluids is always of Ising character but that crossover from nonasymptotic, close-to-classical behavior occurs at scales that may sometimes be experimentally inaccessible \cite{fisher94, weingartner01}. 

 Monte Carlo  computer  simulations provide another useful method of investigating the properties of ionic systems. In the last decade substantial progress has been achieved in this field, with primary effort focused on the coexistence curves \cite{pana92,pana99, orkoulas99}. However, some special attention has also been devoted to the heat capacity  which is significant for elucidating the critical region \cite{fisher94, valleau98, luijten01}. Nevertheless, because of the long-range character of the interactions and the low values of the critical temperatures, which lead to many strongly bound ion pairs \cite{orkoulas99,caillol96}, computer  simulations for finite systems still lack the ability to clearly determine critical exponents and  hence to identify the nature of the criticality.

The success of the renormalization group (RG) approach in describing non-ionic fluids  \cite{fisher74} suggests that it might also be applied to Coulombic criticality. However, to implement an RG treatment, the existence of a  physically well based mean-field theory turns out to be crucial \cite{fisher94}. The simplest model for theoretical investigations of ionic systems is the so called restricted primitive model (RPM), which considers particles of equal sizes and positive and negative charges of equal magnitude. Two main theoretical approaches have emerged. The first employs an extension \cite{fisher94, levin94, levin96, lee97} of the basic Debye-H\"uckel (DH) theory \cite{debye}, developed in the early 20th century for dilute solutions of strong electrolytes. The second approach rests on integral equations for correlation functions, typically employing the Ornstein-Zernike equations in combination with some truncation, as in the mean spherical approximation (MSA) \cite{levin96, stell95,yeh96,ciach01}. Neither of these two approaches has any known independent basis, such as an overall variational principle for the ionic free energy, that might help justify its reliability. However, compared to the values predicted by DH-based treatments, MSA-based theories yield relatively poor agreement with the critical parameters found by current Monte Carlo simulations, namely, in reduced units \cite{fisher94,weingartner01}, $T^*_c \simeq 0.049$ and $\rho^*_c= 0.06$ to  0.085. Careful analysis \cite{fisher94,levin96, zuckerman97}, utilizing thermodynamic energy bounds, etc., also suggests that DH-based theories promise a better description of the critical region of model electrolytes.   

Since the Ising model, which is equivalent to a lattice gas, has played a crucial role in understanding critical phenomena in non-ionic systems, lattice models of electrolytes deserve attention. Although clearly artificial as regards the description of real ionic solutions, which possess continuous rather than discrete spatial symmetry, they are attractive for various reasons. First, by virtue of the lattice character one can incorporate   the behavior of dense phases, at least one of which should be an ordered ionic crystal. Lattice models may also be effective for describing defects in real crystals \cite{walker}. Second, even finely discretized lattice systems present a computational advantage over their continuous-space counterparts in Monte Carlo simulations \cite{orkoulas99}. Last but not the least, discrete-state lattice models facilitate the derivation of equivalent field-theoretical descriptions and, thereby, the  study of the significance of various terms in the effective Hamiltonian. Moreover, Coulomb interactions can be exactly represented in terms of a nearest-neighbor Hamiltonian via the sine-Gordon transformation that yields both low fugacity and high-temperature expansions of the equation of state. (For a recent overview and results see Ref.\cite{caillol01} and references therein.)  

Despite their distinct theoretical advantages, lattice models of electrolytes have not been studied systematically. The aim of the work reported below has been to repair this omission. 

Most of the previous analytical \cite{ciach01,dickman99} and numerical \cite{orkoulas99,ciach01, dickman99, bresme00,almarza01} work on lattice ionic systems has addressed the question of tricriticality and of order-disorder transitions. While the overall density is the order parameter which suffices to reveal gas-liquid critical behavior in ionic solution, the presence of an underlying lattice allows naturally for the appearance of another order parameter. In bipartite lattices, such as the simple cubic (sc) and body-centered cubic (bcc) lattices, ions of opposite charges can distribute unequally between the sublattices, thereby reducing the electrostatic part of the free energy. At the same time, the entropy is also reduced which increase the free energy. This competition leads to the appearance of a phase with long-range order resembling an ionic crystal; second-order phase transitions are then a likely consequence. In the continuum case analogous oscillations appear in the charge-charge correlation functions \cite{lee97} at certain values of density-temperature ratio. However, such ordered phases may turn  out to be thermodynamically so stable, that a gas-liquid phase transition predicted by a continuum theory may not survive in a lattice model: the lattice system  tends to ``solidify''  before forming a ``proper'' liquid. Indeed, this scenario has been observed in numerical studies. On the other hand, the possible presence of both gas-liquid and tricritical points has been predicted theoretically by Ciach and Stell for a model with additional short-range interactions added to the lattice Coulomb forces \cite{ciach01}.

As indicated, we present in this article a study of the simplest, single-site hard core model of lattice ionic system with charges $q_\pm = \pm q$. In Sec. II we describe the basic DH theory on general, $d$-dimensional Bravais lattices. Our analysis focuses on $d=3$ in Sec. III. After presenting the results for pure DH theory, the crucial phenomenon of Bjerrum ion pairing is introduced in Sec. III.B; but, following Fisher and Levin \cite{levin96}, this must be supplemented by explicit dipole-ion solvation effects: see Sec. III.C. Then, in Sec. IV the possibility of sublattice charge ordering is discussed. Unlike previous treatments \cite{dickman99, ciach01}, we account for both electrostatic screening and sublattice ordering in a unified framework. Our conclusions are summarized briefly in Sec. V.

\section{Lattice Debye-H\"uckel  theory in general dimensions}

Our derivation for general $d$-dimensional lattices follows closely the Debye-H\"uckel approach \cite{mcquarrie}. We confine ourselves to the lattice restricted primitive model (LRPM), which consists of oppositely charged ions with charges $q$ and $-q$ which occupy single lattice sites of a $d$-dimensional Bravais lattice. In this simplest model the ions interact only through the electrostatic field and otherwise behave as ideal particles, subject only to on-site exclusion. Thus the total free energy density, which plays the central role in the thermodynamics of the system, can be written $f=f^{Id}+f^{DH}$.  As the overall system must be neutral, the average densities of the positive and negative ions are equal: $\rho_+=\rho_-=\frac{1}{2}\rho_1$. Correspondingly, for the reduced chemical potential and pressure \cite{levin96} we have
\begin{equation} \label{state}
   \bar \mu_+ = \bar \mu_- =\bar \mu_1, \;\;\;\;\; \bar p = \max_{\rho_1} [\bar f + \bar\mu_1\rho_1],
\end{equation}
where $\bar \mu = \mu/ k_{B}T$ and $\bar p= p/ k_{B}T$. The ideal lattice gas contribution to the free energy is,  up to a constant term, given by
\begin{equation} \label{f.id}
    \bar f^{Id}=-\frac{F}{kT V}=-\frac{\rho^*_1}{v_0}\ln\rho_1^* - \frac{1-\rho^*_1}{v_0}\ln(1-\rho^*_1),
\end{equation}
with the corresponding chemical potential
\begin{equation}\label{mu.id}
   \bar \mu_1^{Id} = -\partial \bar f^{Id} / \partial \rho_1 = \ln\rho^*_1-\ln(1-\rho^*_1),
\end{equation}
where $V$ is the total lattice volume while  $\rho^*_1=\rho_1 v_0$ is the reduced dimensionless density of (free) ions and $v_0$ is the volume per site of the lattice. 

Next we determine the contribution to free energy arising from the Coulombic interactions. However, the lattice form  of the potential, which takes into account the discreteness of the space,  should be used. This lattice Coulomb potential will approach  the continuous  $1/r$ potential asymptotically at large distances, but it differs significantly at small distances. We start with  the  linearized  lattice Poisson-Boltzmann equation, which determines the average electrostatic potential  at  point $\mathbf{r}$. Following the standard DH approach \cite{mcquarrie}, we easily find
\begin{equation} \label{linPB}
   \Delta \varphi(\mathbf{r})=\kappa^2 \varphi(\mathbf{r})-(q C_d /D v_0)\delta(r),  
\end{equation}
where $\kappa^2 = C_d\beta \rho_1 q^2 /D $ is the inverse Debye screening length, with $\beta=1/k_{B}T$. The  constant factor $C_d=2\pi^{d/2}/\Gamma(d/2)$ is determined by the dimensionality of the lattice system \cite{levin94}. In this equation we use the lattice Laplacian defined through 
\begin{equation} \label{lat.lap}
   \Delta \varphi(\mathbf{r}) = \frac{2 d}{c_0 a^2}\sum_{nn} \left[\varphi(\mathbf{r}+\mathbf{a})-\varphi(\mathbf{r})\right],
\end{equation}
where $\mathbf{a}$ is a nearest-neighbor vector and the summation runs over all $c_0$ nearest neighbors. The DH equation, (\ref{linPB}), can be easily solved by Fourier transformation yielding
\begin{equation} \label{phiofr}
   \varphi(\mathbf{r})=\frac{C_d q}{D  v_0} \frac{a^2}{2 d} \int_k \frac{e^{i\mathbf{k\cdot r}}}{(x^2+2d)/2d-J(\mathbf{k})},
\end{equation}
with $x=\kappa a$ and $\int_k \equiv (2\pi)^{-d}\int_{-\pi}^{\pi}d^d \mathbf{k}$. The lattice function $J(\mathbf{k})$ is defined by
\begin{equation} \label{functionJ}
   J(\mathbf{k})=\frac{1}{c_0}\sum_{nn}e^{i\mathbf{k\cdot a}}.
\end{equation} 
In the DH approach, we need only the potential felt by an ion fixed at the origin due to all the surrounding ions. Because of the Bravais symmetry, we can find the total potential at the origin by averaging over the nearest-neighbor sites to obtain  
\begin{equation} \label{pot}
\varphi(\mathbf{r}=0)=\frac{1}{c_0}\sum_{nn}\varphi(\mathbf{a}_{nn}).
\end{equation}
Introducing the integrated lattice Green's function via 
\begin{equation} \label{latGreen}
   P(\zeta)=\int_k \frac{1}{1-\zeta J(\mathbf{k})},
\end{equation}
and using (\ref{phiofr}) we obtain
\begin{equation} \label{phiofa}
   \varphi(\mathbf{0})=\frac{C_d q}{D v_0} \frac{a^2}{2d}  \left[P\left(\frac{2d}{x^2+2d}\right)-1\right].
\end{equation}
The potential due to  a single ion in the absence of other ions is obtained simply by setting $x=0$ in (\ref{phiofr}) which yields
\begin{equation} \label{self}
    \varphi_0(\mathbf{0})=\frac{C_d q}{D  v_0} \frac{a^2}{2d} \left[ P(1)-1 \right] .
\end{equation}
Subtracting this expression  from the total potential (\ref{phiofa}) and using (\ref{latGreen}) yields the potential felt by an ion of charge $q_i$ at $\mathbf{r}=0$ due to all the surrounding ions, namely,
\begin{equation}  \label{potential}
   \psi_i=\frac{C_d q_i}{D v_0}\frac{a^2}{2d} \left[ P\left(\frac{2d}{x^2+2d}\right)-P(1) \right].
\end{equation}

The electrostatic part of the free energy can be found by using the Debye charging procedure \cite{mcquarrie} which yields 
\begin{eqnarray}\label{f.el} \nonumber
    \bar f^{El}&=&-\frac{1}{kT V}\sum F^{ion} = -\rho_1 \beta q_i \sum_i \int_0^1 \psi_i(\lambda q) d\lambda \\
    &=& \frac{1}{4d v_0}\left[ x^2 P(1) - \int_0^{x^2}P\left(\frac{2d}{x^2+2d}\right)d(x^2) \right] \label{DH.pure} .
\end{eqnarray}
 Combining the ideal-gas and Debye-H\"{u}ckel terms yields $\bar f^{DH}= \bar f^{Id}+ \bar f^{El}$ and
\begin{equation}\label{mu.pure}
   \bar \mu_1=\ln \rho^*_1-\ln(1-\rho^*_1)-\frac{\pi a^d}{d v_0 T^*}\left[P(1)-P\left(\frac{2d}{x^2+2d}\right)\right] , 
\end{equation} 
with reduced temperature defined by 
\begin{equation} \label{temper}
   T^*=kTDa^{d-2}/q^2 .
\end{equation}
From this we find the pressure for an arbitrary $d$-dimensional Bravais lattice to be
\begin{equation}   \label{pressure.d}
   \bar p v_0 = -\ln(1-\rho^*_1) + \frac{1}{4 d} \left[x^2 P\left(\frac{2d}{x^2+2 d}\right) - \int_0^{x^2}P\left(\frac{2 d}{x^2 +2 d}\right)d(x^2)\right] .
\end{equation}

Eqs.(\ref{f.id}) and (\ref{DH.pure})-(\ref{pressure.d}) give full information about the thermodynamic behavior of a lattice Coulomb  system. In particular, the possibility of phase transitions and criticality can be investigated by analyzing the spinodals,  and the phase coexistence curves may be obtained by the matching  pressure and chemical potential in coexisting phases. Spinodals are  specified by setting the inverse isothermal compressibility $K_T^{-1}$ to zero, so that
\begin{equation} \label{spinodal}
   \frac{1}{\rho_1 k T K_T}= \rho_1 \frac{\partial\bar \mu}{\partial \rho_1}=0 ,  
\end{equation}
which, on taking (\ref{mu.pure}) into account, reduces to  
\begin{equation} \label{spinodal.d}
    T^*_s =\frac{C_d a^d}{2d v_0} \frac{\zeta(1-\zeta) \partial P(\zeta)/\partial \zeta}{2+(1-\zeta)^2 \partial P(\zeta)/\partial \zeta} ,
\end{equation}
with $\zeta=2d/(x^2+2d)$. One can show that when $\rho^*_1$ becomes large (which corresponds to $\zeta\to 0$), one has  $T^*_s \approx c_0 C_d a^d (\rho^*_1)^2/2 d v_0$. Eqs.(\ref{f.el})-(\ref{spinodal.d}) can be used to investigate the phase behavior of electrolytes in any dimension.

We mention briefly here the critical parameters obtained for $d=1$ and $2$. In the one-dimensional case the lattice Green's function gives
\begin{equation}
   P(\zeta)=\frac{1}{\pi}\int_0^\pi \frac{dk}{1-\zeta \cos k}=\frac{1}{\sqrt{1-\zeta^2}} ,
\end{equation}
which yields a spinodal of the form
\begin{equation}
   T^*_s=\frac{2}{x\left[x+(x^2+4)^{3/2}\right]}, \;\;\;\;\;  x=\kappa a.
\end{equation}
This specifies a critical point with parameters 
\begin{equation}
   \rho^*_c=0, \;\;\;\;\;\; T^*_c=\infty,
\end{equation}
in accordance with the general principle that one-dimensional systems do not display phase transitions. However, since $\varphi(r)\propto |r|$ in a one-dimensional system, the DH method fails and describing the one-dimensional ionic lattice model demands a different approach: one may note the continuum analysis \cite{lenard61}. 

For $d=2$ dimensions the lattice Green's functions are given in Ref. \cite{katsura71}. Then for both for  triangular and square lattices the predicted critical parameters are found to be
\begin{equation}
    \rho^*_c=0, \;\;\;\;\;\; T^*_c=1/4,
\end{equation}
precisely, the same values  as for the continuum model \cite{levin94}.

\section{Electrolytes in three dimensions}

\subsection{Pure DH theory}

Let us now examine  $3D$ cubic lattices in more detail. We address three  cases: simple cubic (sc), body centered cubic (bcc) and face centered cubic (fcc); for convenience  their geometrical parameters are listed in Table \ref{table.3dlat}. 
\begin{table}[t] \caption{Lattice parameters} \label{table.3dlat}
  \begin{center} 
  \begin{tabular}{l|c|c|c|c}
     lattice & unit cell   &  nearest  neighbor   & number of nearest & volume per\\ 
             & edge                & distance, $a_{nn}$ & neighbors, $c_0$  & site, $v_0$ \\ \hline 
     sc      & $a_0$           & $a_0$              & 6                 & $a_0^3$      \\
     bcc     & $2a_0$          & $a_0\sqrt{3}$      & 8                 & $4a_0^3$     \\
     fcc     & $2a_0$          & $a_0\sqrt{2}$      & 12                & $2a_0^3$
   \end{tabular}
   \end{center}   
\end{table}
The basic lattice function  $J(\mathbf{k})$, defined in (\ref{functionJ}), is then given by
\begin{eqnarray} \label{j.3}
   J(\mathbf{k}) &=& \textstyle{ \frac{1}{3}}(\cos k_1 + \cos k_2 +\cos k_3) , \quad\quad\quad\quad\quad\quad\quad\quad\quad\quad  \mathrm{sc} , \\
                   &=& \cos k_1\; \cos k_2\; \cos k_3,  \quad\quad\quad\quad\quad\quad\quad\quad\quad\quad\quad\quad\quad  \mathrm{bcc} , \\
                   &=& \textstyle{\frac{1}{3}}(\cos k_1\, \cos k_2 + \cos k_2\, \cos k_3 + \cos k_1\, \cos k_3) , \;\;\,\quad   \mathrm{fcc} ,
\end{eqnarray} 
with $-\pi \le k_1,k_2,k_3 \le \pi$. The corresponding integrated lattice Green's functions can be explicitly calculated using their representation in terms of complete elliptic integrals as shown by Joyce \cite{joyce}. The self-potential of an ion (in the absence of any screening) is then given by  
\begin{eqnarray} \label{self3d}
   (Da/q) \varphi_0(0) &=& \frac{C_d}{v_0} \frac{a^3}{2d} \left[P(1)-1\right] \\
                       &\simeq& 1.082, \quad  1.070 \quad \mathrm{and} \quad  1.021 ,  \nonumber
\end{eqnarray}
for sc, bcc and fcc lattices. This reduced value approaches the exact continuum potential value $1$ as the number of nearest neighbors increases. At low densities the free energy as given by (\ref{f.el}) can be expanded in powers of $x=\kappa a$, which yields 
\begin{eqnarray}
       \bar f^{El}&=&\frac{x^3}{12\pi a^3}(1-0.282x+0.025x^2+...),  \;\;\;\;\;\; \mathrm{sc} , \\
       &=&\frac{x^3}{12\pi a^3}(1-0.286x+0.025x^2+...),  \;\;\;\;\;\;\mathrm{bcc} ,\\
       &=&\frac{x^3}{12\pi a^3}(1-0.296x+0.025x^2+...),  \;\;\;\;\;\;\mathrm{fcc} .
\end{eqnarray}
The leading term precisely reproduces the exact continuum DH result, which, of course, is independent of $a$. The magnitude of the first correction term increases with increasing coordination number; in the hard sphere continuum model it becomes  $0.75$. 

The predicted coexistence curves for the sc, bcc and fcc lattices are shown in Fig. 1, while the critical parameters are listed in Table II. A surprising feature of these coexistence curves is that the liquid density approaches a finite value, $\rho^*_{liq}(0)$, as $T\to 0$ that is substantially smaller than the maximum, close-packing density $\rho_1^*=1$: see Table II.

For comparison, Fig. 1 also displays the predictions of DH theory for the continuum RPM supplemented by hard-core interactions in the free volume approximation with the simple cubic packing limit  \cite{levin96}. Although the critical temperatures decrease slightly as the number of nearest neighbors approaches realistic values of the coordination number, say, $12-14$, as observed in simple liquids, its value for all three lattices remains about 50\% higher than the corresponding continuum value. This is, indeed, a rather general feature of lattice models, which tend to display higher critical temperatures than their continuum-space counterparts. However, the predicted critical densities are quite comparable, decreasing from about 60\% above the continuum value to only 3 or 4 \% higher: see Table III. Clearly, packing considerations play a significant role in the value of $\rho^*_c$.

\begin{figure}
\begin{center}
\unitlength 1in
\begin{picture}(4.5,3.3)
\resizebox{4.5in}{3.3in}{\includegraphics{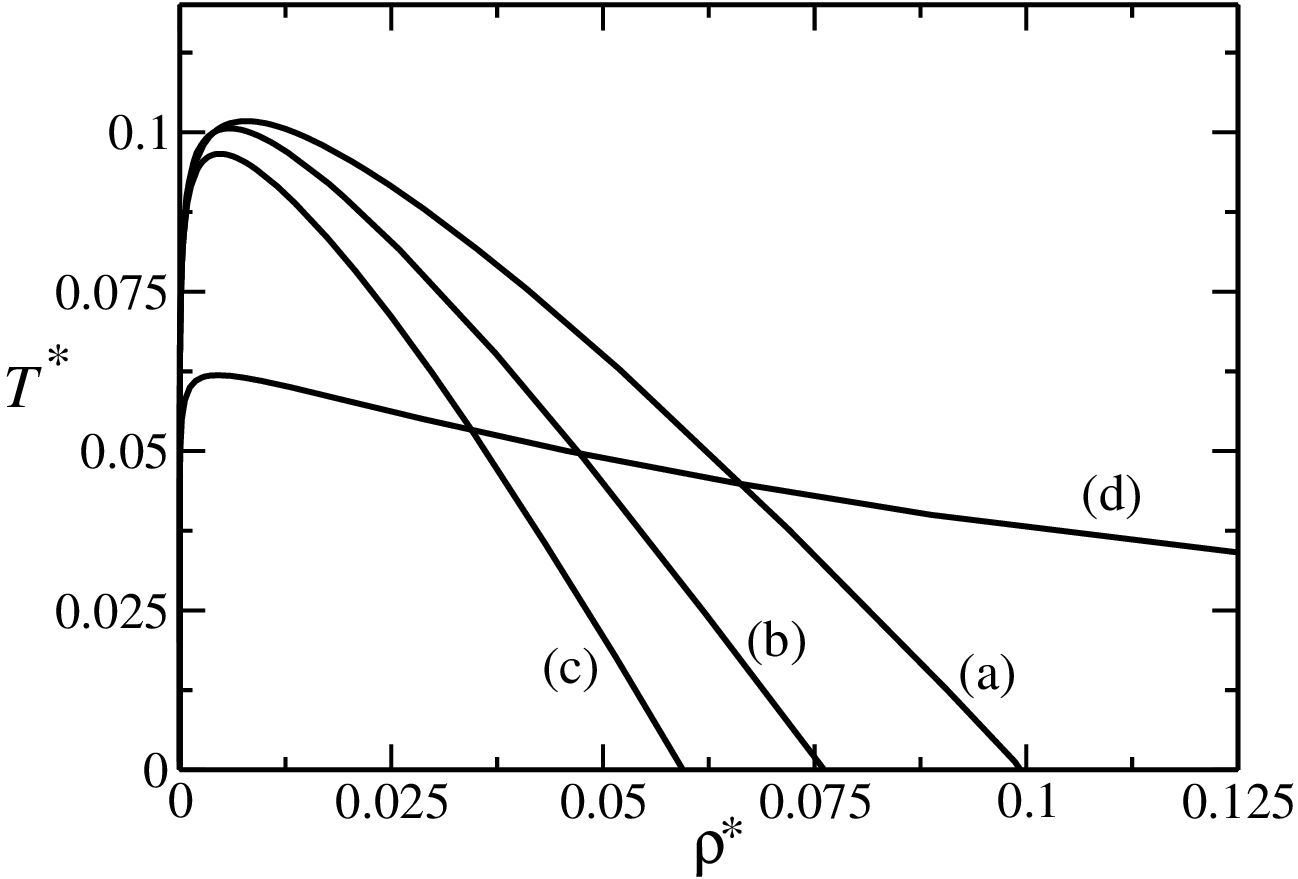}}
\end{picture}
\caption{\label{fig:dh}Coexistence curves for the LRPM predicted by pure DH theory: (a) sc, (b) bcc, (c) fcc, (d) continuum RPM \cite{levin96} with hard-core interactions corresponding to the simple cubic packing limit. }
\end{center}
\end{figure}

\subsection{Bjerrum ion pairing}

Free ions alone are not adequate for treating the low temperature critical region, since positive and negative ions will often combine into strongly bound neutral dimers or Bjerrum pairs \cite{bjerrum}. This process can be treated as a reversible chemical reaction, say, $(+)+(-)~ \rightleftharpoons~ (+,-)$, leading to equilibrium densities of free ions and dipoles, varying with $T$ and $\rho$ \cite{fisher74,levin96}. In a continuum model, however, there arises a serious question as to precisely what configurations are to be considered as bound pairs \cite{levin96, ebeling71, bjerrum}. In practice, this relates directly to the problem of determining the proper association constant $K(T)$. Bjerrum's original approach \cite{bjerrum} was to introduce a temperature-dependent cutoff distance that would represent, in some sense, the size of dipolar pair. Later Ebeling \cite{ebeling71}, using systematic cluster expansions, obtained more elaborate expression for $K(T)$; it turns out, however, that Bjerrum's form is reproduced asymptotically to all orders at low temperatures. But for a lattice system the situation is intrinsically simpler because a clear and acceptable definition of a bound ion pair is two oppositively charged ions occupying neighboring sites. (Pairs separated by further distances, second nearest neighbors, etc., may be regarded as distinct species and could be considered separately, if necessary \cite{levin94}.)
\begin{table}[t] \label{table.crit.DH}
    \caption{Coexistence curve parameters for $3D$ cubic lattices according to pure DH theory; HC denotes the continuum hard sphere system, i.e., the RPM}
    \begin{center}
     \begin{tabular}{l|c|c|c}
       model       & $T^*_c$  & $\rho^*_c$   &  $\rho^*_{liq}(0)$ \\ \hline 
       sc          & \quad 0.101767 \;\;  & \quad 0.007869  \;\;   & \quad 0.0996 \;\; \\
       bcc         & 0.100617 & 0.005908     & 0.0759 \\
       fcc         & 0.096637 & 0.004755     & 0.0596 \\
       HC          & 0.061912 & 0.004582     & 1 
     \end{tabular}
     \end{center}
\end{table} 

Following this convention, we introduce the density $\rho_2$ and chemical potential $\mu_2$ of Bjerrum pairs which we suppose, initially, behave like ideal lattice particles. The condition of chemical equilibrium,  $\mu_2=2\mu_1$, ensures that $\rho^*_1$ and $\rho^*_2$ are interconnected via the law of mass action. To this end, let 
\begin{equation}
     z_1=\Lambda^3_1/(v_0\zeta_1)e^{\bar \mu_1} , \quad \quad\quad\quad  z_2=\Lambda^6_2/(v_0^2\zeta_2)e^{\bar \mu_2}  
\end{equation}
denote activities of free ions and pairs, respectively, where the $\Lambda_i$ denote the de Broglie wavelengths for which we have $\Lambda_+=\Lambda_-=\Lambda_1$ and $\Lambda_1=\Lambda_2$ (see Ref. \cite{levin96}) while $\zeta_1$ and $\zeta_2$ represent the corresponding internal configurational partition functions. In terms of the activities, the law of mass action states $z_2=\frac{1}{4}K z_1^2$, from which follows \cite{levin96}
\begin{equation}
   K(T)=\frac{\zeta_2}{\Lambda_2^6}\left(\frac{\Lambda_1^3}{\zeta_1}\right)^2 =  \zeta_2(T).
\end{equation}
This definition of $K$ as the internal partition function of a dipolar pair leads naturally to the basic expression
\begin{equation}
  K(T)=v_0 \sum_{nn} e^{-\beta q \varphi(a_{nn};T)} = v_0 c_0 e^{-\beta q \varphi(\mathbf{0};T)},
\end{equation}
where $\varphi(\mathbf{0};T)$ is given by (\ref{phiofa}). 

By using the potential-distribution theorem \cite{widom}, we can then write the free ion density as 
\begin{equation}
   \rho_1^* = z_1 e^{ - \Psi/kT}=v_0 \zeta_1 e^{-\bar \mu^{El}_1}e^{-\bar \mu_1  - \Psi/kT}/\Lambda_1^3,
\end{equation}
where  $\Psi$ is the potential of mean-field force and $\bar \mu_1^{El}$ is given by (\ref{mu.pure}) (with $d=3$) since neutral particles do not contribute to the electrostatic interactions. The second exponential factor here accounts for all the non-Coulombic interactions, since the ionic terms are already taken into account by the factor with exponent $\mu_1^{El}$. Hence, only a hard-core factor is required: this may be taken as the probability that a given lattice is empty, namely,  $1-\rho_1^* - 2\rho^*_2$. In total the ionic chemical potential may thus be expressed as
\begin{equation} \label{mu1.dhbj}
   \bar \mu_1 = \ln\left( \frac{\rho^*_1}{1-\rho^*_1 - 2\rho^*_2}\right) + \ln \left(\frac{\Lambda_1^3}{v_0 \zeta_1}\right) + \mu_1^{El} .
\end{equation} 

To obtain a complementary expression for $\bar \mu_2$ we appeal to the Bethe approximation \cite{nagle}. It corresponds to the zeroth-order term in the series expansion of the grand-partition function for dimers with no attractive interactions and yields 
\begin{equation} \label{bethe}
   z_2 v_0=\frac{(2\rho^*_2/c_0) \big[1-(2\rho^*_2/c_0)\big]}{(1-\rho_1^* - 2\rho^*_2)^2} .
\end{equation}
Thence we obtain  
\begin{equation} \label{mu2.dhbj}
    \bar \mu_2 =2\ln\left(\frac{2\rho^*_2}{1-\rho_1^* - 2\rho^*_2}\right) + \ln\left(2\rho^*_2/c_0\right) + \ln\!\big[1-(2\rho^*_2/c_0)\big] - \ln\left(\frac{4 \Lambda^6_2}{\zeta_2 v_0}\right).
\end{equation}
On using  the law of mass action, the Bethe approximation also yields an equation for $\rho^*_2$, namely, 
\begin{equation}
  \frac{(2\rho^*_2/c_0) \big[1-(2\rho^*_2/c_0)\big]}{(1-\rho_1^* - 2\rho^*_2)^2}  = \left(\frac{\rho^*_1}{1-\rho^*_1 - 2\rho^*_2}\right)^2 \frac{c_0}{4}e^{2\mu^{El}_1-\beta q \varphi(\mathbf{0})} .
\end{equation}
Taking into account that the dimer density should increase as the free-ion density increases, we may solve to obtain
\begin{equation} \label{rho2.dhbj}
   \rho^*_2=\frac{c_0}{4}\left[1-\left(1-c_0\rho_1^{*^2}\exp\left\{\frac{2\pi a^3}{3v_0 T^*}\left[P\left(\frac{6}{x^2+6}\right)-1\right]\right\}\right)^{1/2}\right] .
\end{equation}
Since the dimers are neutral, they do not add to the DH interaction energy which retains the form (\ref{DH.pure}). For the total free energy we then have
\begin{equation} \label{f.dhbj}
   \bar f = \bar f^{Id} + \bar f^{El} =2\bar f^{Id}(\textstyle{\frac{1}{2}}\rho_1)+\bar f^{Id}(\rho_2) + \bar f^{El}(\rho_1) ,
\end{equation}
in which we recall that
\begin{equation}
    x=\kappa a = 4\pi a^3\rho_1^* / v_0T^* .
\end{equation}
Now we may note that the free energy density can be found by integration of (\ref{mu1.dhbj}) or (\ref{mu2.dhbj}) with respect to $\rho_1$ or  $\rho_2$, respectively. Comparing the resulting expressions yields  
\begin{equation}  \label{f1.id.dhbj}
    2 \bar f^{Id}_1( {\textstyle\frac{1}{2}} \rho_1) v_0 =- \rho^*_1 \ln\rho^*_1 - (1-\rho_1^* - 2\rho^*_2)\ln(1-\rho_1^* - 2\rho^*_2) -\rho^*_1 \ln\left(\Lambda_1^3 /v_0\right),
\end{equation}
which can  be obtained independently by noting, that the free volume available for an ion is proportional to $1-\rho_1^* - 2\rho^*_2$ (see also Ref. \cite{walker}). In addition we get 
\begin{eqnarray}  \label{f2.dhbj}  \nonumber
   \bar f_2^{Id} v_0 =&-&2\rho^*_2 \ln\rho^*_2 - \rho^*_2\ln(2\rho^*_2/c_0) + (c_0/2) \ln\!\big[1-(2\rho^*_2/c_0)\big]\\ 
                     &-&\rho^*_2 \ln(1-2\rho^*_2/c_0) -\rho^*_2 \ln\left(\Lambda_1^6/\zeta_2 v_0\right)  .
\end{eqnarray}

The equation of state for DH theory with ideal dimers then follows from (\ref{state}). As in the  continuum RPM \cite{levin96}, one finds that the lattice DHBj theory merely superimposes the pressure of an ideal lattice gas of Bjerrum pairs on the DH pressure for the free ions. Additionally, the ideal-gas term for the free ions is changed somewhat, since the hard-core interactions intrinsic to a lattice gas, appear in the entropic contributions to the free energy via the fraction of available lattice sites. However, this does not affect the critical temperature of the liquid-gas transition, since it is primarily the free ions density that governs the phase separation. 

One now finds, just as in the continuum model \cite{levin96}, that the coexistence curves for all the lattices have a banana-like shape: see Fig. 2. This is simply a consequence of the rapid growth of the number of neutral dimers as the temperature increases. Indeed, the overall critical density is predicted to increase by a factor of 5.02 for the sc, 6.38 for the bcc and 10.25 for the fcc lattice, taking the values $\rho^*_{c}=0.03982$, 0.03769, $0.04878$, respectively. Since $\rho^*_{1c}$ does not change when adding ideal dimers, and decreases when the lattice symmetry is enlarged, the fact that  the  overall critical density for fcc lattice is greater than for the bcc and sc lattices is surprising; but, no doubt, the increased coordination number, $c_0$, serves to enhance the formation of dimers.
\begin{figure}
\begin{center}
\unitlength1in
\begin{picture}(4.5,3.5)
\resizebox{4.5in}{3.5in}{\includegraphics{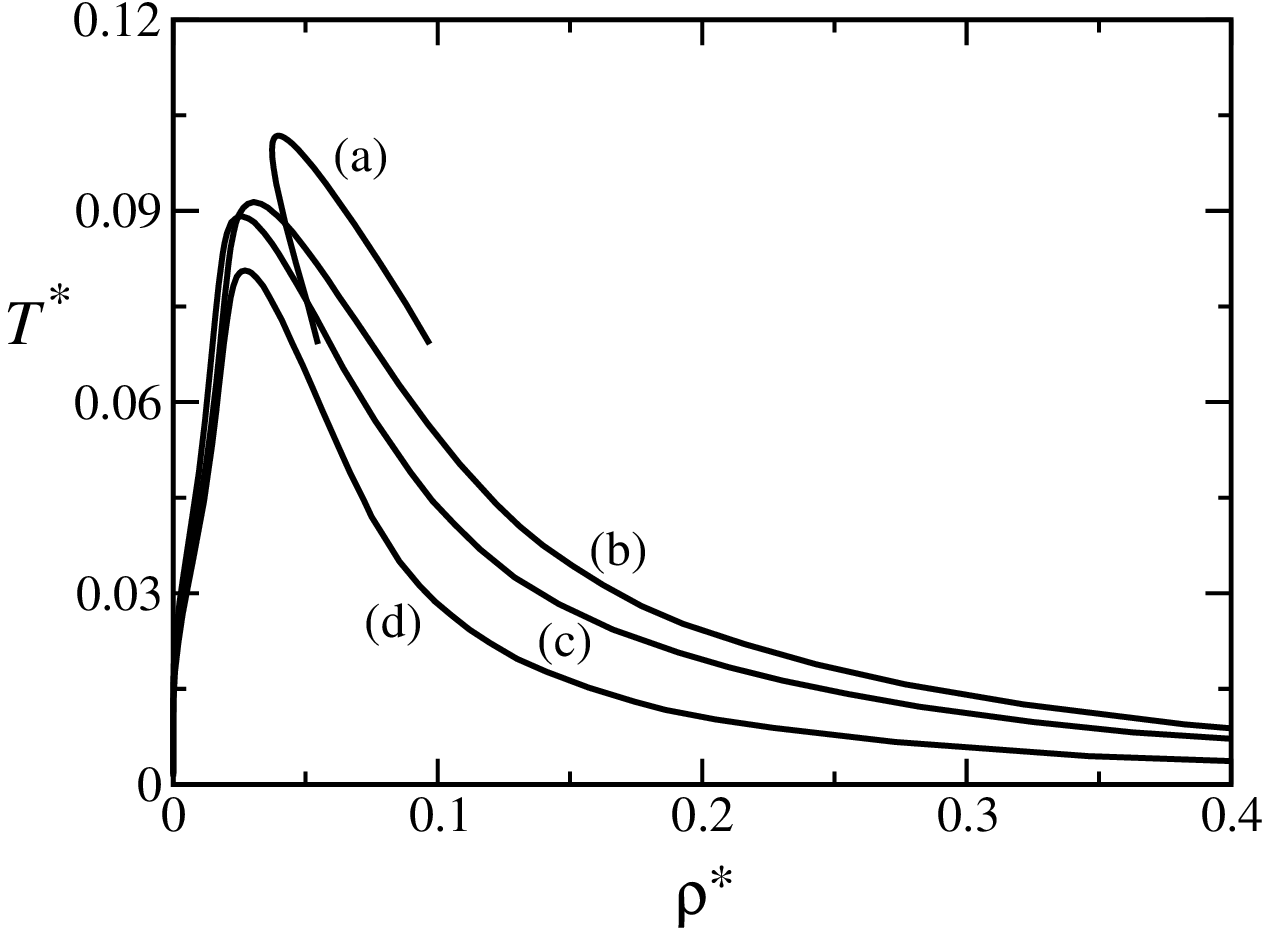}}
\end{picture}
\caption{\label{fig:dhbjdi}Predicted phase diagrams of gas-liquid coexistence: (a) sc lattice with inclusion of Bjerrum pairing alone; for bcc and fcc lattices the coexistence curves have a similar form. Coexistence curves  for (b) sc, (c) bcc, (d) fcc lattices based on the full DHBjDI theory.}
\end{center}
\end{figure}

\subsection{Dipole-ion interactions}

The banana-like shape of the DHBj coexistence curves is clearly unphysical \cite{levin96}. Indeed, as noted by Fisher and Levin \cite{levin96}, the next terms in the expansion of the free energy with respect to density take into account the effects of screening of the bare dipole field of a Bjerrum dimer by the free ions. As shown in Ref. \cite{levin96}, this solvation effect reduces the free energy of an ion pair in the electrolyte. It also eliminates the unphysical banana form of the coexistence curve and produces better agreement between the critical point predictions and the estimates from simulations. 

Proceeding in this direction, the dipole solvation energy $f^{DI}$ can be calculated via the standard DH charging procedure \cite{levin96}. Moreover, for the lattice case it turns out that one can go substantially farther than for continuum-space models. Indeed, under a few very reasonable approximations, we can obtain closed analytical expressions. Consider the positive ion of a dipolar pair.  Instead of (\ref{pot}) we now have
\begin{equation} 
    \varphi_+ (x) = \frac{1}{c_0}\left(\varphi_{-}(x) + \sideset{}{^\prime}\sum_{nn} \varphi(\mathbf{a}_{nn}) \right) ,
\end{equation}
where the prime on the summation means that the site with the negative ion is excluded. Owing to the symmetry, the potentials of the negative and positive ions differ only in sign, the energies being equal. Hence we obtain
\begin{equation} \label{phi+}
    \varphi_+(x) = \frac{1}{c_0+1} \sideset{}{^\prime}\sum_{nn} \varphi(\mathbf{a}_{nn}) ,
\end{equation}
 where the potentials $\varphi(\mathbf{a}_{nn})$ can be calculated using the DH expression (\ref{phiofr}) separately for the contributions arising from the negative and positive ion of each pair. After some algebra the results for the cubic lattices can be written in the general form 
\begin{eqnarray} \label{psi.dhbjdi} 
    \psi_+ &=&\varphi_+(x)-\varphi_+(x=0) \nonumber \\
                                          \\
           &=& \frac{2\pi q a^2}{3 Dv_0}  \frac{1}{c_0+1} \left\{\int_k \frac{1+(c_0-1)J(\mathbf{k})-c_0J^2(\mathbf{k})}{1+x^2/6-J(\mathbf{k})} - 1 \right\} , \nonumber
\end{eqnarray}
with appropriate values of $c_0$ and $J(\mathbf{k})$ for each of the lattices: see Eq.(\ref{j.3}). Utilizing the definition of $P(\zeta)$ in (\ref{latGreen}) enables us to rewrite this in the more convenient form
\begin{equation}
   \psi_+ = \frac{2\pi q a^2}{ 3 D v_0} \frac{1}{c_0+1} \left[{\textstyle \frac{1}{6}} c_0x^2+G(x^2)\right] ,
\end{equation}
in which 
\begin{equation}
   G(x^2)= \frac{x^2[c_0x^2+6(c_0+1)]}{6(x^2+6)}P\left(\frac{6}{x^2+6}\right) .
\end{equation}
Then the DH charging process may be implemented straightforwardly to yield
\begin{equation} \label{f.di}
   \bar f^{DI}= \frac{\pi q^2 a^2 }{3 D v_0^2}\frac{1}{c_0+1} \beta \rho^*_1 \rho^*_2 \left[-\frac{c_0}{12}+\frac{1}{x^2}\int_0^{x^2}G(x^2)d(x^2) \right] ,
\end{equation}
with the corresponding chemical potentials 
\begin{equation} \label{mu1.di} 
  \bar \mu_1^{DI} = \frac{4\pi^2}{3(c_0+1)}\frac{a^6}{v_0^2}\frac{\rho_2^*}{T^{*^2}} \left[\; \frac{c_0}{12} +\frac{1}{x^4}\int_0^{x^2}G(x^2)d(x^2)-  G(x^2)  \right] ,
\end{equation}
\begin{equation} \label{mu2.di}
   \bar \mu_2^{DI}=\frac{\pi}{3(c_0+1)}\frac{a^3}{v_0^2}\frac{1}{T^*}  \left[\frac{c_0}{12}-\frac{1}{x^2}\int_0^{x^2}G(x^2)d(x^2) \right] ,
\end{equation}
and pressure 
\begin{equation} \label{p.di}
   \bar p^{DI}v_0 = \bar f^{DI} v_0 + \bar\mu^{DI}_1 \rho^*_1 + \bar\mu^{DI}_2 \rho^*_2 .
\end{equation}

The only matter not yet taken into account is that, owing to dipole-ion interactions, the excess chemical potential will appear also in the law of mass-action. Since at the densities of interest for criticality we suppose Bjerrum pairs interact only with free ions --- in the continuum-space RPM dipole-dipole interactions appear in the next higher term in the series expansion --- the  Bethe approximation for the dimer activity remains adequate. Thus we obtain an equation, defining implicitly the pair density $\rho^*_2 $ as a function of $\rho^*_1$, namely,
\begin{eqnarray}
  (2\rho^*_2/c_0) \big[1-(2\rho^*_2/c_0)\big]& = &\frac{1}{4} c_0(\rho_1^*)^2 \exp\bigg\{\frac{2\pi a^3}{3v_0T^*}\left[P\left(\frac{6}{x^2+6}\right)-1\right] \bigg. \\
& & +  \bigg. 2 \bar\mu^{DI}_1(\rho^*_1,\rho^*_2)-\bar\mu^{DI}_2(\rho^*_1,\rho^*_2)\bigg\}. \nonumber
\end{eqnarray} 

\begin{table}[t] \label{table.dhbjdi}
    \caption{Critical parameters predicted by the full DHBjDI theory. For comparison, values for the RPM are also given.}
     \begin{center}
     \begin{tabular}{c|c|c}
       model      & $T^*_c$     & $\rho^*_c$     \\ \hline 
       sc lattice         & 0.09666     & 0.03041      \\
       bcc lattice        & 0.08931     & 0.02563      \\
       fcc lattice        & 0.08064     & 0.02708      \\
       RPM: DHBjDI & \quad 0.0554-0.0522 \quad\, & \quad 0.0244-0.0259 \quad      \\
       RPM: simulations \;   &  0.049 &0.06-0.085
     \end{tabular}
     \end{center}
\end{table}

This completes the principal task and allows the construction of coexistence curves: these are shown in Fig. 2. Clearly, in the lattice models the dipole-ion interactions are  also  crucial to repair the unphysical banana-like form produced by Bjerrum association alone. The numerical estimates are presented in Table III. For comparison, the predictions of continuum-space DHBjDI theory and of the RPM simulations are also listed. The predicted lattice critical temperatures are now 1.5-1.65 times greater than the value given by the simulations and the theoretical results of Levin and Fisher \cite{levin96}. The critical densities, however, are quite close to the continuum model predictions, but are significantly lower than the simulations \cite{pana92,pana99, luijten01}.

\section{Sublattice ordering}

So far we have dealt only with an intrinsically low-density picture of the system. Our description of the dense phases, although partially represented by the liquid side of the coexistence curve, has been seriously incomplete. On the other hand, lattice theories provide a particularly natural first approach to studying ordering in solid phases. Indeed, the question of principal interest for us will be the possibility of ordering similar to that observed in ionic crystals. We will, in fact, find that a DH-based theory yields a phase diagram with no gas-liquid criticality but, rather a tricritical point \cite{pana99,dickman99,ciach01}.

\subsection{DH Mean Field Theory}

Let us start by considering a  general $d$-dimensional bipartite lattice, that can be divided into two sublattices of the same form. Suppose, $N_A^+$, $N_A^-$ and $N_B^+$, $N_B^-$ are the numbers of positive and negative ions on sublattice $A$ and $B$, respectively, subject to the neutrality constraint $N_A^++N_A^- = N_B^+ +N_B^-=N/2$. Consider the sublattice with an excess of positive ions (say, ``$A$'' for the definiteness), and define the corresponding order parameter by 
\begin{equation}
 y=\frac{N_A^+ -N_A^-}{N_A^+ +N_A^-} .
\end{equation}
This will have a vanishing mean value in a disordered phase but will be  positive in an appropriately ordered phase. 

The entropic part of the free energy density corresponds to ideal ions and is thus now given by
\begin{equation}
   f^{Id}=-\rho^*_1\ln\rho^*_1-(1-\rho^*)\ln(1-\rho^*_1) -{\textstyle \frac{1}{2}}\rho^*_1 \left[(1+y) \ln(1+y)+(1-y)\ln(1-y)-2\ln2\right] .
\end{equation}
To estimate the electrostatic part of the free energy, the extended  DH approach \cite{lee97} suggests that we begin with an inhomogeneous version of Poisson's equation for the potential at a general site $\mathbf{r}$ due to all the ions when an ion of type $\sigma$ is fixed at the origin: this states 
\begin{equation} \label{eq.or}
  D  \Delta \varphi_\sigma(\mathbf{r})=-C_d \sum_\tau q_\tau\rho_\tau(\mathbf{r})g_{\tau,\sigma}(\mathbf{r})+C_d q_\sigma \delta(\mathbf{r}),
\end{equation}
where $\rho_\tau(\mathbf{r})=\rho^*_\tau(\mathbf{r})/v_0$ is the bulk density of ions of species $\tau$ while $g_{\tau,\sigma}(\mathbf{r})$ is the ion-ion correlation function. Approximating the correlation functions by simple Boltzmann factors and then linearizing provides a DH equation. 

However, we must now allow for an overall nonzero charge density on each sublattice given by
\begin{equation} \label{rho.ex}
   n_A = \frac{\sum_\tau N^\tau_A q_\tau}{N/2} = \rho^*_1 y q, \quad\quad  n_B = \frac{\sum_\tau N^\tau_B q_\tau}{N/2} = -\rho^*_1 y q .
\end{equation}
These charge densities generate an additional ``background'' potential, $\Phi(\mathbf{r})$, which does not contribute to the correlation functions since it is independent of what type of charge is placed at the origin. For sublattice $A$ we thus have a linearized Poisson-Boltzmann or DH equation
\begin{equation}
   \Delta \Phi(\mathbf{r}_A)=-\frac{C_d}{D}\rho^*_1 y q = - \frac{\kappa^2 y}{\beta q} .
\end{equation}
Recalling the definition of the lattice Laplacian (\ref{lat.lap}), and taking into account the symmetry between the sublattices we have
\begin{equation} \label{fiab}   
    \Phi(\mathbf{r}_A)=-\Phi(\mathbf{r}_B=\mathbf{r}_A+\mathbf{a}_{nn}) ,
\end{equation}
and conclude that the background potentials are 
\begin{equation} \label{phi.ex.1}
   \Phi(\mathbf{r}_A)= - \Phi(\mathbf{r}_B) =\frac{x^2 y}{4d \beta q}  .
\end{equation}

If we now put  $\tilde \varphi_A = \varphi_A-\Phi(\mathbf{r}_A)$ and accept the approximation $g_{\tau,\sigma}(\mathbf{r}) \simeq \exp(-\beta\tilde\varphi)$ in (\ref{eq.or}), the equation for the local induced potential $\tilde \varphi_A$ reduces to the lattice version of pure DH theory (\ref{linPB}). This reflects the electrostatic superposition principle, i.e., the total potential is simply the sum of a ``background'' potential due to nonzero average charge density and the DH screening potential so that
\begin{equation} \label{phi.a}
   \varphi(\mathbf{r})=\Phi(\mathbf{r})+\frac{C_d q}{D v_0} \frac{a^2}{2 d} \int_k \frac{e^{i\mathbf{k\cdot r}}}{1+ x^2/2d-J(\mathbf{k})} .
\end{equation}
Now, following the DH approach combined with a mean-field description of the ordering, we find the potential $\psi$ due to all ions except the one fixed at the origin  to be 
\begin{eqnarray} 
     \varphi_A(\mathbf{r}_A=0)& = & \frac{1}{c_0}\sum_{nn}\left[\Phi(\mathbf{a}_{nn})+\varphi^{DH}(\mathbf{a}_{nn})\right] ,\\
     \psi_A & =& \varphi_A(\mathbf{0})-\varphi_A(\mathbf{0})|_{x=0} ,
\end{eqnarray}
which, on taking into acount (\ref{fiab}) and (\ref{phi.ex.1}),  yields  
\begin{equation} \label{psi.ex}
   \psi_A =  -\frac{x^2 y}{4d \beta q}+\psi^{DH} ,
\end{equation}
with $\psi^{DH}$ given by the same expression as $\psi_i$ in (\ref{potential}). 

Finally, the DH charging process gives the total free energy density (of both sublattices) as $\bar f = \bar f^{Ord} + \bar f^{DH}$ where $ \bar f^{DH}=\bar f^{Id}+\bar f^{El}$ follows from (\ref{f.id}) and (\ref{f.el}), while
\begin{equation}
    \bar f^{Ord} = \frac{C_d a^d}{8 d v_0^2}\frac{(\rho_1^*)^2 y^2}{T^*} - \frac{\rho^*_1}{2 v_0}\left[(1+y)
\ln(1+y)+(1-y)\ln(1-y)-2\ln2 \right] .
\end{equation}
 Note that this result implies that the electrostatic part of the ordering energy is negative ($\bar f=-F/kTV$)  as it should be since it describes the interactions between charges of opposite signs. The ordering term  also yields additions to the chemical potential and pressure, namely, 
\begin{equation}
    \bar \mu^{Ord} = -\frac{C_d a^d}{4 d v_0}\frac{\rho^{*}_1 y^2}{T^*} + {\textstyle \frac{1}{2}}\left[(1+y)
\ln(1+y)+(1-y)\ln(1-y)-2\ln2\right],
\end{equation}
\begin{equation}
    \bar p^{Ord} = C_d a^d \rho^{*^2}_1 y^2 /8 d v_0^2 T^* .
\end{equation}
 
Now the possibility of sublattice ordering  is explored by seeking minima of $\bar f^{Ord}$ with $y \ne 0$. This leads to
\begin{equation}
    \frac{C_d a^d \rho^*_1 y}{2 d v_0} = \ln\left(\frac{1+y}{1-y}\right) .
\end{equation}
Expanding for small $y$ in the standard way yields the solution
\begin{equation} \label{y.order}
    y \approx \sqrt{3}\left(\frac{\rho^*_1}{\rho^*_\lambda}-1\right)^{1/2},
\end{equation}
in which the $\lambda$-line, along which second order phase transitions occur, is given by  
\begin{equation}\label{lambda}
     \rho^*_\lambda (T) = (4 d v_0 /C_d a^d) T^* .
\end{equation}
The simplest way to find the anticipated tricritical point is to consider the intersection of the spinodal with the $\lambda$-line \cite{dickman99}. This can be found by computing $\partial \bar p(\rho^*_1,y) / \partial \rho^*_1$ with $y$ defined by (\ref{y.order}), equating to zero and setting $\rho^*_1=\rho^*_\lambda(T)$ after taking the derivative. A more readily justifiable, but also somewhat more sophisticated procedure is to study the stability matrix for the free energy, which is now a function of two order parameters, namely, $y$ and $\rho^*_1$. Both methods lead to the same equation for tricritical point, which reads
\begin{equation} \label{tricrit}
   \frac{4 d}{\rho^*_{tr}}\frac{\partial P}{\partial x^2}\!\!\left(\frac{2d}{x^2+2d}\right)\!\bigg|_{x^2=4d} + \frac{1}{1-\rho^*_{tr}} -\frac{3}{2} = 0 ,
\end{equation} 
where $P(\zeta)$ is the lattice Green's function (\ref{latGreen}). 

\subsection{Results and Discussion}

In $d=3$ dimensions simple cubic and body centered cubic lattices are bipartite and sublattice ordering is possible. Note, that the calculations presented above did not use any extra  properties of lattice symmetry. Indeed, the electrostatic interaction energy of the ``charged'' sublattices depends only on the excess charge density, that is on $y$, and is thus the same for both lattices. This is also true as regards the entropy of sublattice ordering. We find that the tricritical point parameters are
\begin{eqnarray} \label{tri}
   T_{tri}^* &=& 0.3822\; (\mathrm{sc}), \quad 0.4865 \;(\mathrm{bcc}), \\
   \rho_{tri}^* &=& 0.3649 \;(\mathrm{sc}), \quad 0.3576\; (\mathrm{bcc}),
\end{eqnarray}
while the $\lambda$-lines may be written 
\begin{equation}
   T_{\lambda}^*/ \rho^*_{\lambda} = {\textstyle \frac{1}{3}} \pi \simeq 1.047 \; (\mathrm{sc}), \quad {\textstyle \frac{1}{4}}\pi \sqrt{3} \simeq 1.360 \; (\mathrm{bcc}) .
\end{equation}
The full predicted phase diagrams are presented in Fig. 3.   

\begin{figure}
\begin{center}
\unitlength1in
\begin{picture}(4.5,3.5)
\resizebox{4.5in}{3.5in}{\includegraphics{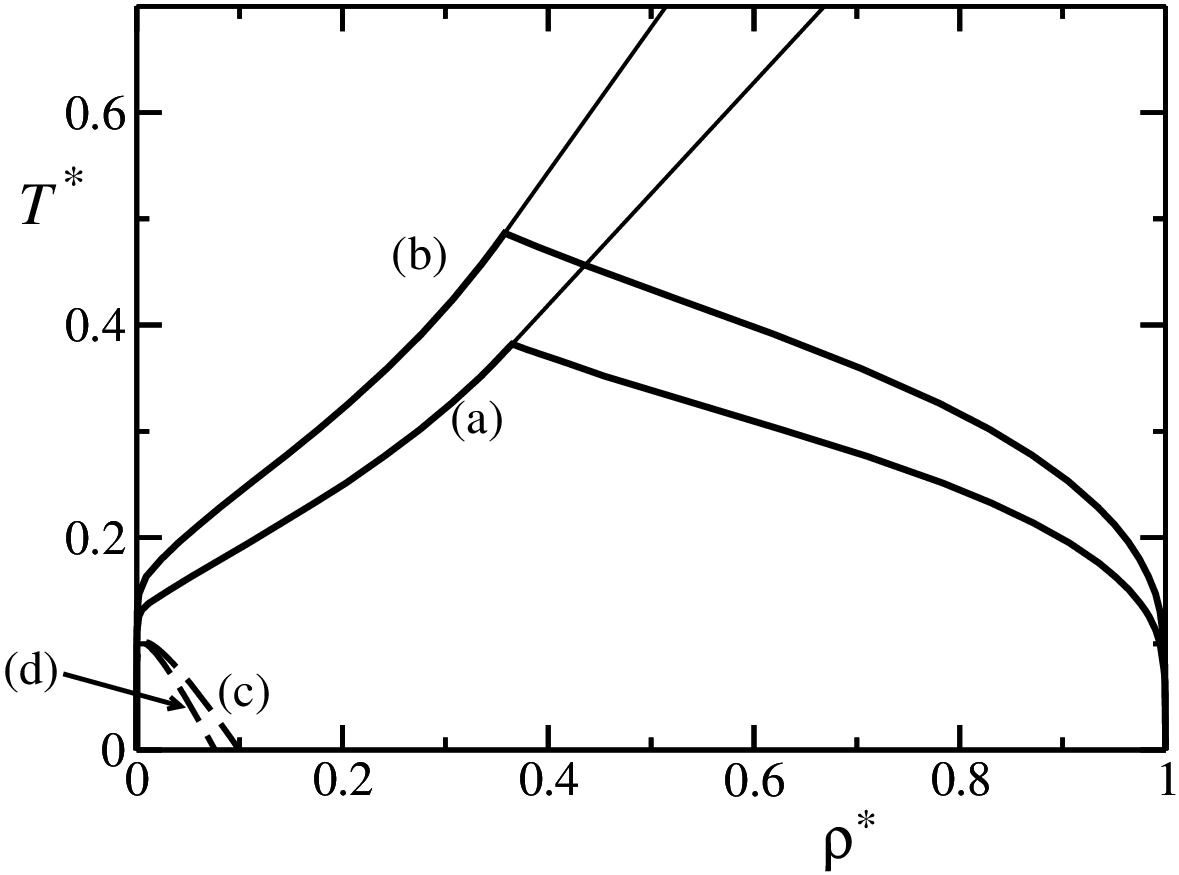}}
\end{picture}
\caption{\label{fig:order} Phase diagrams for ionic lattice systems with sublattice ordering: (a) sc, (b) bcc. Also shown by broken lines are the gas-liquid separation curves predicted by pure DH theory for (c) the  sc and (d) the bcc lattice. }
\end{center}
\end{figure}

It is instructive to compare our results with previous simulations for the sc lattice \cite{dickman99,orkoulas99}. These yield $T^*_{tr}\simeq 0.14 - 0.15$ which is only about 40\% of our theoretical estimates (\ref{tri}). On the other hand, for the tricritical density the higher estimate $\rho_{tri}^* \simeq 0.48$ of Ref. \cite{orkoulas99} is probably more reliable than $\rho_{tri}^* \simeq 0.38$ of Ref. \cite{dickman99} (which compares rather well with our theoretical values), since the former simulations used larger lattice sizes and computed more points on the coexistence curve. It must also be noted, however, that both these simulations employed the discretized continuum or $1/r$ Coulombic potential in place of the lattice form we have used.

Fitting a straight line at low $T$ to the $\lambda$-line data of Stell and Dickman \cite{dickman99} yields a slope $  T_{\lambda}^*/ \rho^*_{\lambda}  \simeq 0.6 $ which may be compared with our value of 1.047. (For another comparison one might note that the generalized DH theory for the continuum RPM \cite{lee97} predicts damped charge-density oscillations setting in on a locus   $T_{K}^*/ \rho^*_{K} \simeq 0.3$ while undamped oscillations are predicted beyond the locus $T_{X}^*/ \rho^*_{X} \simeq 9$.)

Also of interest is the value of $T_{\lambda}^*(\rho^*)$ at close packing, i.e., $\rho^* =1$. For the sc lattice our treatment predicts $T_{\lambda}^*(1)\simeq 1.047$ which, by virtue of the mean-field character of the theory, is likely to be a significant overestimate. Indeed, extrapolation of the Dickman-Stell data suggests $T_{\lambda}^*(1)\simeq 0.6$ while Almarza and Enciso \cite{almarza01} obtain $T_{\lambda}^*(1)\simeq 0.515$; but, again, these simulations employ a discretized  $1/r$  potential. Our analysis indicates higher values of $T^*_{tri}$ and $T^*_{\lambda}(1)$ for the bcc lattice than for the sc. But, perhaps, surprisingly, this is just the opposite of what Almarza and Enciso find \cite{almarza01}. In addition to using the lattice Coulomb potential, our treatment at this point has neglected the formation of Bjerrum ion pairs. In fact, it seems quite feasible to include pairing in the theory along with allowance for sublattice ordering (since the ions of a dimer pair will reside on complementary sublattices). It is possible that this improvement of the theory will result in lower transition temperatures for the bcc lattice relative to the sc lattice. 

Previous theoretical discussions of the sc lattice RPM have been presented by Stell and colleagues \cite{dickman99, ciach01}. In an initial mean field approach \cite{dickman99}, the long-range Coulomb potential (taken in discretized $1/r$ form) was first reduced to an effective nearest-neighbor interaction. The value of the tricritical temperature, $T_{tri}^* =2$, obtained by direct mean-field lattice calculations, was then scaled down by a factor derived by comparing  energy magnitudes. This approach suggested $T_{tri}^* \simeq 0.3$ and $\rho^*_{tri}=\frac{1}{3}$, the latter value being merely a consequence of using a nearest-neighbor mean-field approximation. More recently, Ciach and Stell \cite{ciach01} have adopted a single-ion lattice potential, as, in fact, given by (\ref{phiofr}) and (\ref{phiofa}). This corresponds more closely to our treatment but they entirely neglect the cooperative screening which must occur and which is included in our DH-based treatment. (Note that at higher densities the screening effectively takes place via ``holes'' in the ordered or close-to ordered lattice charge configurations). The new treatment \cite{ciach01} reproduces $\rho^*_{tri}=\frac{1}{3}$ (for the previous reasons) but gives $T_{tri}^* \simeq 0.6$, which is worse than the previous result as compared with the simulations; however, no energy  rescaling is now performed.

As we have seen, both by our own theoretical analysis and through the simulations, the sc and bcc pure Coulomb lattice systems display \textit{no} gas-liquid phase separation as such. Indeed, in Fig. 3 we have plotted the coexistence curves for the two lattices that are predicted by pure lattice DH theory with no allowance for the possibility of sublattice ordering. Evidently,  these coexistence curves lie entirely within the two-phase coexistence region of dilute disordered vapor and the high density ordered ``crystal''. Consequently, the order-disorder transition entirely suppresses gas-liquid criticality in the simplest discrete ionic system with only \textit{single-site} hard cores. If, instead, the hard-core repulsions extend over more lattice sites --- or, equivalently, if a finer level of spatial discretization is employed (as, of course, is more-realistic for continuum systems) --- then, as revealed by simulations \cite{pana99,orkoulas99}, a normal gas-liquid transition and critical point is restored. At the same time, ordered, crystal-like phases appear only at relatively higher densities as characteristic of real solids.

While a DH (or, even, a DHBjDI) theory might be attempted for a more finely discretized model, the clearly evident complications do \textit{not} make this a promising prospect. On the other hand, by adding to a purely ionic lattice system strong short-range attractive potentials (say, designed to represent neutral solvent properties \cite{ciach01}), more elaborate phase diagrams can be anticipated. Indeed, by approximations that again neglect all screening effects, systems displaying \textit{both} a tricritical point and a normal critical point have been obtained \cite{ciach01}. Our more complete treatment could readily be extended in the same spirit.

\section{Conclusions}

By solving exactly lattice versions of the usual Debye-H\"uckel equation, we have derived closed expressions for the free energy of general, $d$-dimensional ionic lattice systems with single-site hard core repulsions. For $d\ge3$, gas-liquid transitions are predicted at low temperatures and densities. As in the corresponding DH-based theory  for the continuum restricted primitive model \cite{levin96}, improvement of the theory at low temperatures demands both allowance for $(+,-)$ ion  pairing, to form nearest-neighbor dipolar dimers, {\it and} the solvation of the resulting dipolar pairs by the residual free ions. The predicted critical temperatures for the sc, bcc and fcc lattices in $d=3$ dimensions then lie  $60 - 70 \%$ higher than given by continuum DH-based theories; but the critical densities are relatively closer. These results accord with the general  tendency of lattice theories to overestimate the stability of the corresponding low-temperature continuum phases.

At higher densities in a lattice theory it is imperative to allow for sublattice ordering of the positive and negative ions. By introducing an appropriate order parameter we have extended analysis to treat general, $d$-dimensional bipartite lattices at a \textit{combined} Debye-H\"uckel and mean-field ordering level. Our unified theory yields, in an accord with recent lattice simulations, a complete suppressiion of gas-liquid phase separation and criticality by order-disorder transitions that occur at higher temperatures. At high densities and temperatures a classical second-order $\lambda$-line is predicted; but this terminates at a \textit{tricritical} point at a density, for the sc and bcc lattices, $\rho^*_{tri}/\rho^*_{max}\simeq 0.36$ and a temperature, $T^*_{tri}/T^*_{max}\simeq 0.4 - 0.5$. At lower temperatures the first-order transition is from an exponentially dilute vapor to an almost close-packed ordered ionic lattice.

Our treatment can be extended in various directions. Indeed, there are preliminary indications that by considering strongly anisotropic three-dimensional lattices, gas-liquid separation may be restored, possibly, together with distinct order-disorder transitions. It is relevant to note in this connection that DH theory for continuum ionic systems predicts \textit{ increasing} values of gas-liquid critical temperatures when the dimensionality is decreased \cite{levin94}. Thus lattice anisotropy might mimic lower dimensionality.

Although the direct applicability of our results to ionic systems is clearly limited, we feel the approach developed in Sec. IV may prove helpful in describing the behavior of defects in ionic crystals \cite{walker}. Furthermore, lattice simulations that employ the true lattice Coulombic potential are desirable and might cast some light on the role of short-range interactions and geometric constraints in strongly coupled ionic systems.

\begin{acknowledgments}
The support of the National Science Foundation (through Grant No. CHE 99-81772 to M.E.F.) is gratefully acknowledged. A.B.K. also acknowledges the support of the Camille and Henry Dreyfus New Faculty Awards Program (under Grant No. NF-00-056).
\end{acknowledgments}


\end{document}